\documentclass[conference]{IEEEtran}
\usepackage{graphicx}
\usepackage{amssymb}
\usepackage{amsthm}
\usepackage{amsmath}
\usepackage{caption2}
\usepackage{subfigure}
\usepackage{mathtools}
\usepackage[utf8]{inputenc}

\ifCLASSINFOpdf
\else
\fi

\begin{document}
%
\title{MISO Broadcast Channel with Imperfect and (Un)matched CSIT in the Frequency Domain: DoF Region and Transmission Strategies}


\author{\IEEEauthorblockN{Chenxi Hao and Bruno Clerckx}
\IEEEauthorblockA{Communication and Signal Processing Group, Department of Electrical and Electronic Engineering\\
Imperial College London, United Kingdom\\
Email: \{chenxi.hao10,b.clerckx\}@imperial.ac.uk}}

\maketitle

\begin{abstract}
In this contribution, we focus on a frequency domain two-user Multiple-Input-Single-Output Broadcast Channel (MISO BC) where the transmitter has imperfect and (un)matched Channel State Information (CSI) of the two users in two subbands.  We provide an upper-bound to the Degrees-of-Freedom (DoF) region, which is tight compared to the state of the art. By decomposing the subbands into subchannels according to the CSI feedback qualities, we interpret the DoF region as the weighted-sum of that in each subchannel. Moreover, we study the sum \emph{DoF} loss when employing sub-optimal schemes, namely Frequency Division Multiple Access (FDMA), Zero-Forcing Beamforming (ZFBF) and the $S_3^{3/2}$ scheme proposed by Tandon et al. The results show that by switching among the sub-optimal strategies, we can obtain at least $80\%$ and $66.7\%$ of the optimal sum \emph{DoF} performance for the unmatched and matched CSIT scenario respectively. \footnote{This work was partially supported by the Seventh Framework Programme for Research of the European Commission under grant number HARP-318489.}
\end{abstract}


\IEEEpeerreviewmaketitle

\section{Introduction}
Transmitter side channel state information (CSIT) is crucial to the \emph{DoF} performance in downlink BC, but the CSIT in practice is subject to latency and inaccuracy. Since Maddah-Ali and Tse have showed the usefulness of the delayed CSIT \cite{Tse10}, many researches have investigated the \emph{DoF} region in time domain BC with imperfect instantaneous and stale CSIT \cite{Ges12}\cite{Gou12}\cite{Chen12b}\cite{Tandon12}\cite{Elia13}. However, in practical systems like Long Term Evolution (LTE), the system performance loss is primarily due to CSI measurement and feedback inaccuracy rather than delay. Therefore, assuming the CSI arrives at the transmitter instantaneously, we are interested in the frequency domain BC where the CSI is measured and reported to the transmitter on a per-subband basis. Due to frequency selectivity, constraints on uplink overhead and user distribution in the cell, the quality of CSI reported to the transmitter varies across users and subbands.

\begin{figure}[t]
\renewcommand{\captionfont}{\small}
\captionstyle{center}
\centering \subfigure[Unmatched]{
                \centering
                \includegraphics[width=0.2\textwidth,height=1.25cm]{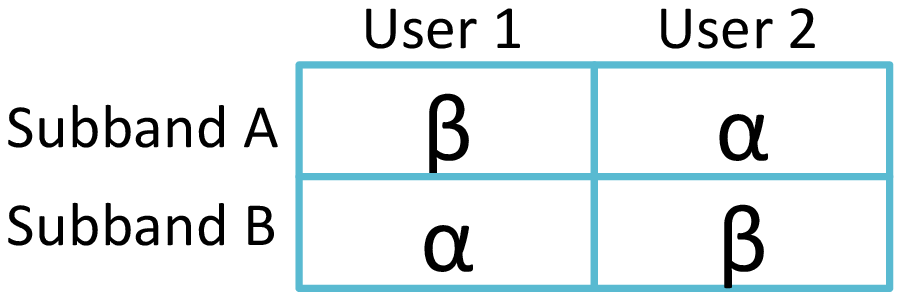}
                \label{fig:unmatched}
        }
\subfigure[Matched]{
                \centering
                \includegraphics[width=0.2\textwidth,height=1.25cm]{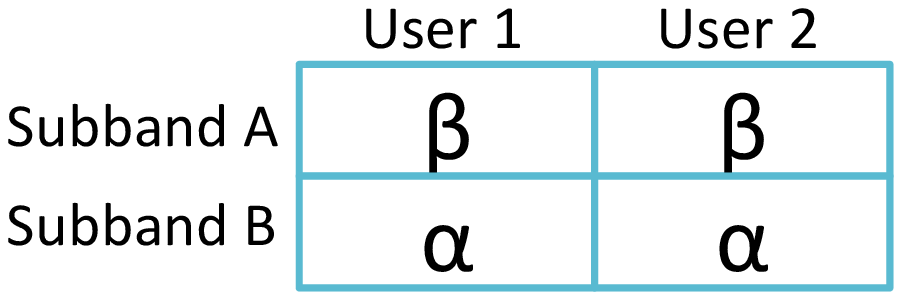}
                \label{fig:matched}
        }
\caption{Two-subband based frequency correlated BC.}\label{fig:scene}
\end{figure}

The work in \cite{Tandon12} has solved the problem when two scheduled users report their CSI on two different subbands (alternating between $I_1I_2{=}NP$ and $PN$\footnote{$I_i$ is the CSIT state of user $i$, it is perfect (P), delayed (D) or none (N).}) by proposing the $S_3^{3/2}$ scheme, achieving optimal sum \emph{DoF} $\frac{3}{2}$. But what if the feedback is imperfect? Literature \cite{icc13freq} was the first work investigating this issue. A novel transmission strategy integrating Maddah-Ali and Tse (MAT) scheme, ZFBF and FDMA is proposed considering a specific two-subband based scenario shown in Figure \ref{fig:unmatched}. However, the \emph{DoF} region found in \cite{icc13freq} is in fact suboptimal and has been improved recently by the scheme proposed in \cite{Elia13}, inspired by the $S_3^{3/2}$ scheme.
\subsection{Main Contributions}
In this paper, we first continue the study in \cite{icc13freq} and \cite{Elia13} by giving a converse in Section \ref{OB}, showing the optimality of the achievable scheme in \cite{Elia13} for the unmatched CSIT. The optimal bound and achievable scheme for the scenario with matched CSIT (see Figure \ref{fig:matched}) are also addressed. Besides, we provide a weighted-sum interpretation of the \emph{DoF} region.

Second, we analyze the achievability of the schemes proposed in \cite{icc13freq} and \cite{Elia13} in Section \ref{dofloss}. The origins of the \emph{DoF} loss in \cite{icc13freq} is clarified.

Third, in Section \ref{switch}, rather than applying a complicated optimal strategy in both unmatched or matched cases, we switch among FDMA, ZFBF and $S_3^{3/2}$ in order to achieve a certain percentage of the optimal sum \emph{DoF} performance. Interestingly, the results show that the optimal scheme can be replaced by the suboptimal switching strategy if we aim at achieving at least $80\%$ and $66.7\%$ of the optimal performance in the unmatched and matched scenario, respectively.

\subsection{Notations}
Bold lower letters stand for vectors whereas a symbol not in bold font represents a scalar. $\left({\cdot}\right)^T$ and $\left({\cdot}\right)^H$ represent the transpose and conjugate transpose of a matrix or vector respectively. $\mathbf{h}^\bot$ denotes the orthogonal space of the channel vector $\mathbf{h}$. $\mathcal{E}\left[{\cdot}\right]$ refers to the expectation of a random variable, vector or matrix. $\parallel{\cdot}\parallel$ is the norm of a vector. $f\left(P\right){\sim}{P^{B}}$ corresponds to $\lim \limits_{P{\to}{\infty}}\frac{{\log}f\left(P\right)}{{\log}P}{=}B$, where $P$ is the SNR throughout the paper and logarithms are in base $2$. For a $V_j$ that is a function of the index $j$, we denote $V_{j_1}^{j_2}$ as the set $\{V_{j_1}{,}V_{j_1{+}1}{,}\cdots{,}V_{j_2}\}$ if $j_1{\leq}j_2$. Otherwise, $V_{j_1}^{j_2}$ is an empty set. 

\subsection{System Model}
\subsubsection{Frequency domain two-user MISO BC}
We consider one 2-antenna transmitter and two single-antenna users. The transmit signal is denoted as $\mathbf{s}_j$, subject to a per-subband power constraint $\mathcal{E}\big[\left\|\mathbf{s}_j\right\|^2\big]{\sim}P$. The observations at user 1 and 2, $y_j$ and $z_j$ respectively, are given by
\begin{equation}
y_j{=}\mathbf{h}_j^H\mathbf{s}_j{+}\epsilon_{j1},\quad
z_j{=}\mathbf{g}_j^H\mathbf{s}_j{+}\epsilon_{j2},\quad j{=}A{,}B\label{eq:model}
\end{equation}
where $\epsilon_{j1}$ and $\epsilon_{j2}$ are unit power AWGN noise. $\mathbf{h}_A$ and $\mathbf{h}_B$, both with $\mathbf{I}_2$ covariance matrix, are the CSI of user 1 in \emph{subband $A$} and $B$ respectively. $\mathbf{g}_A$ and $\mathbf{g}_B$ are those of user 2. The CSI are i.i.d across users and subbands.
\subsubsection{CSI Feedback Model}
Classically, in Frequency Division Duplexing (FDD), each user estimates their CSI in the specified subband using pilot and the estimated CSI is quantized and reported to the transmitter via a rate-limited link. In Time Division Duplexing (TDD), CSI is measured on the uplink and used in the downlink assuming channel reciprocity. We assume a general setup (valid for both FDD and TDD) where the transmitter obtains the CSI instantaneously, but with imperfectness, due to the estimation error and/or finite rate in the feedback link.

Denoting the imperfect CSI in \emph{subband $j$} as $\hat{\mathbf{h}}_j$ and $\hat{\mathbf{g}}_j$, the CSI of user 1 and user 2 can be respectively modeled as $\mathbf{h}_j{=}\hat{\mathbf{h}}_j{+}\tilde{\mathbf{h}}_j$ and $\mathbf{g}_j{=}\hat{\mathbf{g}}_j{+}\tilde{\mathbf{g}}_j$, where $\tilde{\mathbf{h}}_j$ and $\tilde{\mathbf{g}}_j$ are the error vectors, respectively with the covariance matrix $\mathbb{E}[\tilde{\mathbf{h}}_j\tilde{\mathbf{h}}_j^H]{=}\sigma_{j1}^2\mathbf{I}_2$ and $\mathbb{E}[\tilde{\mathbf{g}}_j\tilde{\mathbf{g}}_j^H]{=}\sigma_{j2}^2\mathbf{I}_2$. $\hat{\mathbf{h}}_j$ and $\hat{\mathbf{g}}_j$ are respectively independent of $\tilde{\mathbf{h}}_j$ and $\tilde{\mathbf{g}}_j$. The norm of $\hat{\mathbf{h}}_j$ and $\hat{\mathbf{g}}_j$ scale as $P^0$ at infinite SNR.

To investigate the impact of the imperfect CSIT on the \emph{DoF} region, we assume that the variance of the error exponentially scales with SNR, namely $\sigma_{j1}^2{\sim}P^{-a_j}$ and $\sigma_{j2}^2{\sim}P^{-b_j}$. $a_j$ and $b_j$ are respectively interpreted as the quality of CSIT of user 1 and user 2 in subband $j$, given as follows
\begin{equation}
a_j{=}\lim_{P{\to}\infty}{-}\frac{{\log}\sigma_{j1}^2}{{\log}P}{,}\quad b_j{=}\lim_{P{\to}\infty}{-}\frac{{\log}\sigma_{j2}^2}{{\log}P},
\end{equation}
where $P$ is the SNR throughout the paper since the variance of the AWGN noise has been normalized.

Figure \ref{fig:unmatched} shows the scenario with unmatched (alternating) CSIT, where $a_1{=}b_2{=}\beta$ and $a_2{=}b_1{=}\alpha$. Without loss of generality, we assume that $\beta{\geq}\alpha$. Figure \ref{fig:matched} illustrates the scenario with matched CSIT, namely $a_1{=}b_1{=}\beta$ and $a_2{=}b_2{=}\alpha$.

$\beta$ and $\alpha$ vary within the range of $\left[0{,}1\right]$. $\beta{=}1$ (resp. $\alpha{=}1$) is equivalent to perfect CSIT because the full \emph{DoF} region can be achieved by simply doing ZFBF. $\beta{=}0$ (resp. $\alpha{=}0$) means that the variance of the CSI error scales as $P^0$, such that the imperfect CSIT cannot benefit the \emph{DoF} when doing ZFBF.

\subsubsection{DoF Definition}
The \emph{DoF} is defined on a per-channel-use basis as
\begin{equation}
d_k \triangleq \lim_{P\to\infty} \frac{R_k}{r\log P},\quad k=1,2, \label{eq:dof_def}
\end{equation}
where $R_k$ is the rate achieved by user $k$ over $r$ channel uses.

\section{Outer-Bound of the DoF region}\label{OB}
\newtheorem{mytheorem}{Theorem}

\begin{mytheorem} \label{outerbound}
The outer-bound of the DoF region in the frequency correlated BC with imperfect CSIT (for both unmatched and matched scenario) is specified by
\begin{equation}
d_1+d_2\leq1+\frac{\beta+\alpha}{2},\quad
d_1\leq1,\quad d_2\leq1.\label{eq:bd}
\end{equation}
\end{mytheorem}

\subsection{Proof of Theorem \ref{outerbound}}
Let us first revisit the converse in previous literatures. In \cite{HJSV09}, the \emph{DoF} region in the BC without CSIT is upper-bounded by considering one user's observation is degraded compared to the other's. In the BC with delayed CSIT \cite{Tse10}\cite{Ges12}\cite{Gou12}, the outer-bound is obtained through the genie-aided model where one user's observation is provided to the other, thus establishing a physically degraded BC to remove the delayed CSIT.

However, in this contribution, those methods are not adopted since the transmitter does not have delayed CSIT and the BC with imperfect CSIT cannot be simply considered as a degraded BC. Instead, we follow the assumption in \cite{Lapidoth}: We first consider that user 2 knows the message intended to user 1, which leads to an outer-bound denoted by $\mathbb{D}_1$; Then by assuming that user 1 knows user 2's desired message, we can have another region $\mathbb{D}_2$. The final \emph{DoF} outer-bounds results from the intersection of them, i.e. $\mathbb{D}{=}\mathbb{D}_1{\Cap}\mathbb{D}_2$. This assumption is somehow consistent with the outer-bound given by Theorem 5 in \cite{Marton}, which is used to find a tight upper bound on the weighted sum rate in vector Gaussian BC (Section 4.1, \cite{extremal}).

We assume that the transmission lasts for $n$ subbands ($n{\to}\infty$), half of which are \emph{subband $A$} and the rest are \emph{subband $B$}. The set of the observations of user 1 and user 2 from \emph{subband} $j_1$ to $j_2$ are defined as $Y_{j_1}^{j_2}$ and $Z_{j_1}^{j_2}$ respectively. The set of the imperfect CSI of user 1 and 2 are respectively represented as $\hat{\mathcal{H}}_1^n$ and $\hat{\mathcal{G}}_1^n$ and they are known at both transmitter side and receiver side. $\tilde{\mathcal{H}}_1^n$ and $\tilde{\mathcal{G}}_1^n$ are the set of errors vectors which are only available at user 1 and user 2 respectively. The transmit signal $\mathbf{s}_j$ in subband $j$ is any sequence of $(2^{nR_1}{,}2^{nR_2}{,}n)$ as a function of $W_1$ (user 1's messages), $W_2$ (user 2's messages) and $\hat{\mathcal{H}}_1^n{,}\hat{\mathcal{G}}_1^n$.

Considering that user 2 knows $W_1$, we derive $\mathbb{D}_1$ as follows
\begin{align}
nR_1{\leq}&I(W_1;Y_1^n|\hat{\mathcal{H}}_1^n{,}\hat{\mathcal{G}}_1^n{,}\tilde{\mathcal{H}}_1^n)\\
{=}&\underbrace{h(Y_1^n|\hat{\mathcal{H}}_1^n{,}\hat{\mathcal{G}}_1^n{,}\tilde{\mathcal{H}}_1^n)}_{{\leq}h(Y_1^n){\leq}n{\log}P}{-}
h(Y_1^n|W_1{,}\hat{\mathcal{H}}_1^n{,}\hat{\mathcal{G}}_1^n{,}\tilde{\mathcal{H}}_1^n)\nonumber\\
{\leq}&n{\log}P{-}h(Y_1^n|W_1{,}\hat{\mathcal{H}}_1^n{,}\hat{\mathcal{G}}_1^n{,}\tilde{\mathcal{H}}_1^n),\label{eq:R11}\\
nR_2{\leq}&I(W_2;Z_1^n|\hat{\mathcal{H}}_1^n{,}\hat{\mathcal{G}}_1^n{,}\tilde{\mathcal{G}}_1^n{,}W_1)\\
{=}&h(Z_1^n|\hat{\mathcal{H}}_1^n{,}\hat{\mathcal{G}}_1^n{,}\tilde{\mathcal{G}}_1^n{,}W_1){-}
\underbrace{h(Z_1^n|\hat{\mathcal{H}}_1^n{,}\hat{\mathcal{G}}_1^n{,}\tilde{\mathcal{G}}_1^n{,}W_1{,}W_2)}_{{\leq}no({\log}P)}\nonumber\\
{\leq}&h(Z_1^n|\hat{\mathcal{H}}_1^n{,}\hat{\mathcal{G}}_1^n{,}\tilde{\mathcal{G}}_1^n{,}W_1).\label{eq:R21}
\end{align}
Hence,
\begin{align}
n(R_1{+}R_2){\leq}&n{\log}P{+}h(Z_1^n|\hat{\mathcal{H}}_1^n{,}\hat{\mathcal{G}}_1^n{,}\tilde{\mathcal{G}}_1^n{,}W_1)\nonumber\\
&{-}h(Y_1^n|\hat{\mathcal{H}}_1^n{,}\hat{\mathcal{G}}_1^n{,}\tilde{\mathcal{H}}_1^n{,}W_1)\nonumber
\end{align}
\begin{align}
{=}&n{\log}P{+}h(Z_1^n|\hat{\mathcal{H}}_1^n{,}\hat{\mathcal{G}}_1^n{,}\tilde{\mathcal{G}}_1^n{,}\tilde{\mathcal{H}}_1^n{,}W_1)\nonumber\\
&{-}h(Y_1^n|\hat{\mathcal{H}}_1^n{,}\hat{\mathcal{G}}_1^n{,}\tilde{\mathcal{G}}_1^n{,}\tilde{\mathcal{H}}_1^n{,}W_1).\label{eq:Rs2}
\end{align}
\eqref{eq:Rs2} follows the fact that $Z_1^n$ is independent of $\tilde{\mathcal{H}}_1^n$ conditioned on $\{\hat{\mathcal{H}}_1^n{,}\hat{\mathcal{G}}_1^n{,}\tilde{\mathcal{G}}_1^n{,}W_1\}$, and $Y_1^n$ is independent of $\tilde{\mathcal{G}}_1^n$ conditioned on $\{\hat{\mathcal{H}}_1^n{,}\hat{\mathcal{G}}_1^n{,}\tilde{\mathcal{H}}_1^n{,}W_1\}$. For convenience, we denote $\Omega{\triangleq}\{\hat{\mathcal{H}}_1^n{,}\hat{\mathcal{G}}_1^n{,}\tilde{\mathcal{G}}_1^n{,}\tilde{\mathcal{H}}_1^n{,}W_1\}$. Consequently,
\begin{align}
n(R_1{+}R_2){=}&n{\log}P{+}h(Z_1^n|\Omega){-}h(Y_1^n|\Omega)\nonumber\\
{=}&n{\log}P{+}\sum_{j{=}1}^{n}\{h(Z_j|\Omega{,}Y_1^{j{-}1}{,}Z_{j{+}1}^n)\nonumber\\
&{-}h(Y_j|\Omega{,}Y_1^{j{-}1}{,}Z_{j{+}1}^n)\}.\label{eq:lemma1}
\end{align}
\eqref{eq:lemma1} is similar to equation (44) in \cite{Lapidoth}. We provide the derivation in the Appendix.

In the following, we introduce a new notation as
\begin{eqnarray}
&\{T_j{,}\tilde{\mathcal{Q}}\}{\triangleq}\{\Omega{,}Y_1^{j{-}1}{,}Z_{j{+}1}^n\},&\nonumber\\ &T_j{\triangleq}\{Y_1^{j{-}1}{,}Z_{j{+}1}^n{,}W_1{,}\hat{\mathcal{H}}_1^n{,}\hat{\mathcal{G}}_1^n\}{,}\quad \tilde{\mathcal{Q}}{\triangleq}\{\tilde{\mathcal{H}}_1^n{,}\tilde{\mathcal{G}}_1^n\}.&\nonumber
\end{eqnarray}

Next, we aim at maximizing each term in the summation of \eqref{eq:lemma1} following the footsteps in \cite{Ges12}. We note
\begin{multline}
h(Z_j|T_j{,}\tilde{\mathcal{Q}}){-}h(Y_j|T_j{,}\tilde{\mathcal{Q}})\\
{\leq}\max_{P_{T_j}P_{\mathbf{s}_j|T_j}}\{h(Z_j|T_j{,}\tilde{\mathcal{Q}}){-}h(Y_j|T_j{,}\tilde{\mathcal{Q}})\},\label{eq:max1}
\end{multline}
where the maximizations are taken over all the possible joint distributions of $P(T_j{,}\mathbf{s}_j)$. We write
\begin{align}
\eqref{eq:max1}{\leq}&\max_{P_{T_j}}\mathcal{E}_{T_j}\{\max_{P_{\mathbf{s}_j|T_j}}h(Z_j|T_j{=}T^*{,}\tilde{\mathcal{Q}}){-}h(Y_j|T_j{=}T^*{,}\tilde{\mathcal{Q}})\}\nonumber\\
{=}&\max_{P_{T_j}}\mathcal{E}_{T_j}\{\max_{P_{\mathbf{s}_j|T_j}}\mathcal{E}_{\tilde{\mathcal{Q}}|T_j}
[h(Z_j|T_j{=}T^*{,}\tilde{\mathcal{Q}}{=}\tilde{\mathcal{Q}}^*)\nonumber\\
&{-}h(Y_j|T_j{=}T^*{,}\tilde{\mathcal{Q}}{=}\tilde{\mathcal{Q}}^*)]\}\nonumber\\
{=}&\max_{P_{T_j}}\mathcal{E}_{T_j}\{\max_{P_{\mathbf{s}_j|T_j}}\mathcal{E}_{\tilde{\mathcal{Q}}}
[h(\mathbf{g}_j^H\mathbf{s}_j{+}\epsilon_{j2}|T_j{=}T^*)\nonumber\\&{-}h(\mathbf{h}_j^H\mathbf{s}_j{+}\epsilon_{j1}|T_j{=}T^*)]\}\nonumber\\
{=}&\max_{P_{T_j}}\mathcal{E}_{T_j}\{\max_{\mathbf{K}{\succeq}\mathbf{0}{,}0{\leq}tr(\mathbf{K}){\leq}P}\mathcal{E}_{\tilde{\mathcal{Q}}}
[h(\mathbf{g}_j^H\mathbf{s}_j{+}\epsilon_{j2}|T_j{=}T^*)\nonumber\\&{-}h(\mathbf{h}_j^H\mathbf{s}_j{+}\epsilon_{j1}|T_j{=}T^*)]\}\nonumber\\
{\leq}&\max_{\hat{\mathcal{Q}}}\mathcal{E}_{\hat{\mathcal{Q}}}\{\max_{\mathbf{K}{\succeq}\mathbf{0}{,}0{\leq}tr(\mathbf{K}){\leq}P}
\mathcal{E}_{\tilde{\mathcal{Q}}}
\left[{\log}(1{+}\mathbf{g}_j^H\mathbf{K}\mathbf{g}_j)\right.\nonumber\\
&\left.{-}{\log}(1{+}\mathbf{h}_j^H\mathbf{K}\mathbf{h}_j)\right]\},\label{eq:extremal}
\end{align}
where $\mathcal{Q}{=}\hat{\mathcal{Q}}{+}\tilde{\mathcal{Q}}$ with $\hat{\mathcal{Q}}{\triangleq}\{\hat{\mathcal{H}}_1^n{,}\hat{\mathcal{G}}_1^n\}$ is the channel state of both users and $\mathbf{K}$ is the covariance matrix of $\mathbf{s}_j$. \eqref{eq:extremal} is derived according to the fact 1) $\mathbf{s}_j{\to}T_j{\to}\mathbf{g}_j$ forms a Markov chain so that $\mathbf{g}_j$ is independent of $\mathbf{s}_j$ conditioned on $T_j$; 2) A Gaussian distributed $\mathbf{s}_j$ conditioned on $T_j$ is the optimal solution to the maximization of the weighted difference in \eqref{eq:extremal}, based on the proof of Corollary 6 in \cite{extremal}.

Using Lemma 1 in \cite{Ges12}, we can respectively upper- and lower-bound the first and second terms in \eqref{eq:extremal} as
\begin{align}
\mathcal{E}_{\tilde{\mathcal{Q}}}{\log}(1{+}\mathbf{g}_j^H\mathbf{K}\mathbf{g}_j){\leq}&
{\log}(1{+}\lambda_1\mathcal{E}[||\hat{\mathbf{g}}_j||^2]){+}O(1),\label{eq:upper}\\
\mathcal{E}_{\tilde{\mathcal{Q}}}{\log}(1{+}\mathbf{h}_j^H\mathbf{K}\mathbf{h}_j){\geq}&
{\log}(1{+}e^{-\gamma}\lambda_1\mathcal{E}[||\tilde{\mathbf{h}}_j||^2]){+}O(1),\label{eq:lower}
\end{align}
where $\gamma$ is constant, $\lambda_1$ is the largest eigen-value of the covariance matrix $\mathbf{K}$. Substituting the terms in \eqref{eq:extremal} with \eqref{eq:upper} and \eqref{eq:lower}, we can upper-bound \eqref{eq:lemma1} by
\begin{equation}
n(R_1{+}R_2){\leq}n{\log}P{+}\sum_{j{=}1}^n
{\log}\frac{1{+}\lambda_1\mathcal{E}[||\hat{\mathbf{g}}_{j}||^2]}{1{+}e^{-\gamma}\lambda_1\mathcal{E}[||\tilde{\mathbf{h}}_{j}||^2]}.\label{ea:Rs}
\end{equation}
As \emph{subband $A$} and $B$ respectively take half of the $n$ subbands, replacing $\mathcal{E}[||\hat{\mathbf{g}}_{j}||^2]$, $\mathcal{E}[||\tilde{\mathbf{h}}_{j}||^2]$ with the corresponding values, we obtain
\begin{equation}
\mathbb{D}_1:d_1{+}d_2{\leq}1{+}\frac{\alpha{+}\beta}{2},\label{eq:D1}
\end{equation}

Switching the role of each user, the same formula is obtained for $\mathbb{D}_2$. \eqref{eq:bd} holds for both unmatched and matched case. $\hfill\Box$

\begin{figure}[t]
\renewcommand{\captionfont}{\small}
\centering
\includegraphics[height=1.25cm,width=8cm]{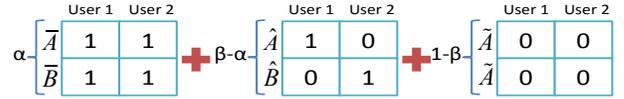}
\caption{Channel decomposition for the unmatched case.}\label{fig:decomp}
\end{figure}
\subsection{A Weighted-Sum DoF Interpretation}
In this part, we decompose the channel in each subband by making use of the intuition that the imperfect CSIT with error variance $P^{-\alpha}$ can be considered as perfect for $\alpha$ ($0{\leq}\alpha{\leq}1$) channel use (i.e. the transmit power is reduced to $\mathcal{E}[||\mathbf{s}||^2]{\leq}P^\alpha$). We can see this by simply sending one private message per user using ZFBF precoding and with power $P^{\alpha}$. Since $\mathcal{E}[|\mathbf{h}_j^H\hat{\mathbf{h}}_j^\bot|^2]{\sim}P^{-\alpha}$ and $\mathcal{E}[|\mathbf{g}_j^H\hat{\mathbf{g}}_j^\bot|^2]{\sim}P^{-\alpha}$, both users can recover their private message only subject to noise. Therefore, the rate $\alpha{\log}P$ is achieved per user. As only $\alpha$ channel has been used, full \emph{DoF} region is obtained according to \eqref{eq:dof_def}. This is in fact a generalization of the fact that  full \emph{DoF} region can be obtained if the error in CSIT is scaled as $SNR^{-1}$ \cite{Ges12}.

Therefore, we decompose the subbands into subchannels as follows (see Figure \ref{fig:decomp}):
\begin{itemize}
  \item $\tilde{A}$, $\tilde{B}$: no CSIT, each with channel use $1{-}\beta$;
  \item $\hat{A}$ ($\hat{B}$): perfect CSIT of user 1 (2), with channel use $\beta{-}\alpha$;
  \item $\bar{A}$, $\bar{B}$: perfect CSIT of both users, with channel use $\alpha$.
\end{itemize}

The \emph{DoF} region in \emph{subbands $A$} and $B$ can be obtained as the weighted-sum of the regions in each subchannel. 

\emph{Subchannel $\tilde{A}$} and $\tilde{B}$ can be categorized as the BC with no CSIT, whose \emph{DoF} region has been studied in \cite{HJSV09}. The outer-bound (denoted as $\tilde{\mathcal{D}}$) is given by
\begin{equation}
\mathcal{D}^{\tilde{A}}=\mathcal{D}^{\tilde{B}}=\tilde{\mathcal{D}}: d_1+d_2\leq1. \label{eq:Dtilde}
\end{equation}

\emph{Subchannel $\bar{A}$} and $\bar{B}$ are the BC with perfect CSIT of both users, the outer-bound is expressed (via a notation $\bar{\mathcal{D}}$) as
\begin{equation}
\mathcal{D}^{\bar{A}}=\mathcal{D}^{\bar{B}}=\bar{\mathcal{D}}: d_1\leq 1,d_2\leq1.\label{eq:Dbar}
\end{equation}

However, \emph{subchannel $\hat{A}$} and $\hat{B}$ have an alternating CSIT setting with two states \cite{Tandon12}: $I_1I_2{=}PN$ and $I_1I_2{=}NP$. The optimal \emph{DoF} region has been found in \cite{Tandon12} as
\begin{align}
(\mathcal{D}^{\hat{A}}{+}\mathcal{D}^{\hat{B}})/2=\hat{\mathcal{D}}:& d_1+d_2\leq1.5,d_1\leq 1,d_2\leq1.\label{eq:Dhat}
\end{align}

Consequently, by combining \eqref{eq:Dtilde}, \eqref{eq:Dbar}, \eqref{eq:Dhat}, we can obtain a weighted-sum representation of the \emph{DoF} region as
\begin{equation}
\mathcal{D}_u=(1-\beta)\tilde{\mathcal{D}}+(\beta-\alpha)\hat{\mathcal{D}}+\alpha\bar{\mathcal{D}}.\label{eq:w_sum}
\end{equation}

Similarly, \emph{subband $A$} and $B$ with matched CSIT can be decomposed as
\begin{itemize}
  \item $\tilde{A}$, $\tilde{B}$: no CSIT, each with channel use $1{-}\beta$ and $1{-}\alpha$;
  \item $\bar{A}$, $\bar{B}$: perfect CSIT of both users, with channel use $\beta$ and $\alpha$ respectively.
\end{itemize}
The weighted sum form of the outer-bound $\mathcal{D}_m$ is given by
\begin{equation}
\mathcal{D}_{m}=(1-\frac{\beta+\alpha}{2})\tilde{\mathcal{D}}+\frac{\beta+\alpha}{2}\bar{\mathcal{D}}.\label{eq:Dmatched}
\end{equation}

\begin{figure}[t]
\renewcommand{\captionfont}{\small}
\centering
\includegraphics[height=6cm,width=8cm]{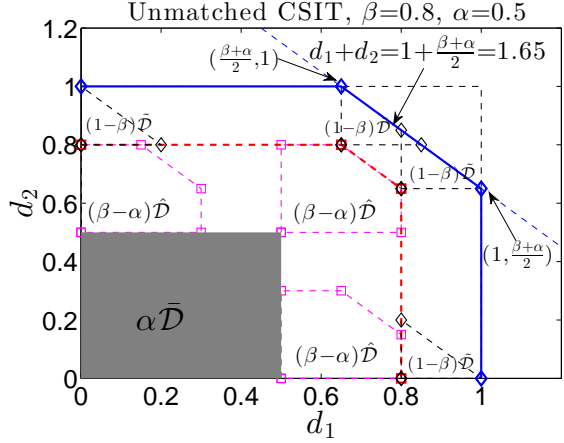}
\caption{Composing $\mathcal{D}_u$ with $\beta{=}0.8$ and $\alpha{=}0.5$}\label{fig:Du}
\end{figure}

\begin{figure}[t]
\renewcommand{\captionfont}{\small}
\centering
\includegraphics[height=6cm,width=8cm]{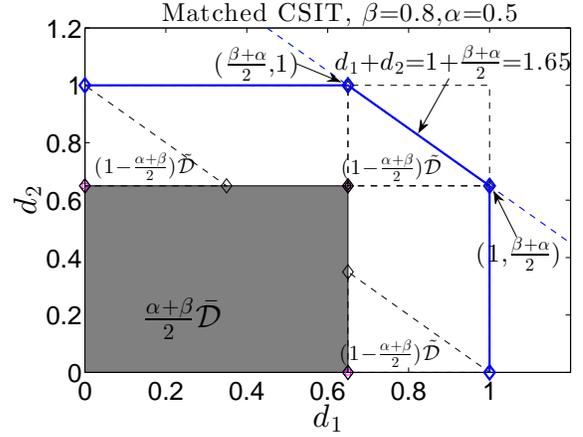}
\caption{Composing $\mathcal{D}_m$ with $\beta{=}0.8$ and $\alpha{=}0.5$}\label{fig:Dm}
\end{figure}

Figure \ref{fig:Du} and \ref{fig:Dm} respectively illustrate the composition of $\mathcal{D}_u$ and $\mathcal{D}_m$. In Figure \ref{fig:Du}, the grey square area depicts the region $\alpha\bar{\mathcal{D}}$. All the valid points inside $\alpha\bar{\mathcal{D}}$ are expanded to a magenta polygon representing $(\beta{-}\alpha)\hat{\mathcal{D}}$. This expansion results in the bound shown by the dashed red curve with square points. Then, every point on this bound is further expanded to a black triangle area referring to $(1{-}\beta)\tilde{\mathcal{D}}$. Outlining all the expanded area, we can obtain $\mathcal{D}_u$ bounded by the solid blue curve with diamond points. It results in inequalities \eqref{eq:bd}. $\mathcal{D}_m$ is made up similarly and illustrated in Figure \ref{fig:Dm}, resulting in \eqref{eq:bd} as well.

The outer-bound shown in Figure \ref{fig:Du} is consistent with the achievable region in \cite{Elia13}, therefore showing that the bound in Theorem \ref{outerbound} is the optimal \emph{DoF} region. The outer-bound illustrated in Figure \ref{fig:Dm} is also an optimal bound, its achievability will be discussed in Section \ref{match_ach}.

\emph{Remark:} The imperfect CSIT setting can be viewed as the alternating CSIT configuration in \cite{Tandon12}, when the weight in front of each term is interpreted as the fraction of the state PP, NP/PN or NN. Also, the value $\frac{\alpha{+}\beta}{2}$ in \eqref{eq:bd} stands for the average quality of CSIT of a user, corresponding to the parameter $\lambda_P$ in \emph{Remark 1} and \emph{2} in \cite{Tandon12}, which represents the fraction of channel use when the CSIT of a user is available.

\section{Achievability Analysis}\label{dofloss}
In this section, we aim at identifying the shortness of the scheme in \cite{icc13freq} in the unmatched case by comparing it with \cite{Elia13} and identifying the optimal scheme for the matched case.
\subsection{Unmatched Case: Revisiting the Optimal Scheme in \cite{Elia13}}\label{unmatch_ach}
The transmit signals in \emph{subband $A$} and $B$ are expressed as
\begin{align}
\mathbf{s}_A &= \left[x_{c,A},0\right]^T{+}
[\hat{\mathbf{g}}_A^\bot,\hat{\mathbf{g}}_A][u_A,u_0]^T{+}\hat{\mathbf{h}}_A^\bot{v_A},\label{eq:sA}\\
\mathbf{s}_B &= \left[x_{c,B},0\right]^T{+}
[\hat{\mathbf{h}}_B^\bot,\hat{\mathbf{h}}_B][v_B,u_0]^T{+}\hat{\mathbf{g}}_B^\bot{u_B}.\label{eq:sB}
\end{align}
$x_{c,A}$ and $x_{c,B}$ are common messages that should be decoded by both users (but could be intended to user 1 and user 2 respectively or exclusively to user 1 or user 2). $u_A$, $u_0$ and $u_B$ are symbols sent to user 1, while $v_A$ and $v_B$ are symbols to user 2. The rate and power allocation are shown in Table \ref{tab:power_rate}, resulting in the received signals at each user ($y_A$ and $y_B$ for user 1 and $z_A$ and $z_B$ for user 2) as
\begin{align}
y_A&{=}\underbrace{h_{A{,}1}^*x_{c,A}}_P{+}\underbrace{\mathbf{h}_A^H\hat{\mathbf{g}}_A^\bot u_A}_{P^\alpha}{+}\underbrace{\mathbf{h}_A^H\hat{\mathbf{g}}_Au_0}_{P^\beta}
{+}\underbrace{\mathbf{h}_A^H\hat{\mathbf{h}}_A^\bot{v_A}}_{P^0},\label{eq:ya}\\
z_A&{=}\underbrace{g_{A{,}1}^*x_{c,A}}_P{+}\underbrace{\mathbf{g}_A^H\hat{\mathbf{g}}_A^\bot u_A}_{P^0}{+}\underbrace{\mathbf{g}_A^H\hat{\mathbf{g}}_Au_0}_{P^\beta}{+}
\underbrace{\mathbf{g}_A^H\hat{\mathbf{h}}_A^\bot{v_A}}_{P^\beta}\!,\\
y_B&{=}\underbrace{h_{B{,}1}^*x_{c,B}}_P{+}\underbrace{\mathbf{h}_B^H\hat{\mathbf{h}}_B^\bot v_B}_{P^0}{+}\underbrace{\mathbf{h}_B^H\hat{\mathbf{h}}_Bu_0}_{P^\beta}{+}
\underbrace{\mathbf{h}_B^H\hat{\mathbf{g}}_B^\bot{u_B}}_{P^\beta},\\
z_B&{=}\underbrace{g_{B{,}1}^*x_{c,B}}_P{+}\underbrace{\mathbf{g}_B^H\hat{\mathbf{h}}_B^\bot v_B}_{P^\alpha}{+}\underbrace{\mathbf{g}_B^H\hat{\mathbf{h}}_Bu_0}_{P^\beta}
{+}\underbrace{\mathbf{g}_B^H\hat{\mathbf{g}}_B^\bot{u_B}}_{P^0},\label{eq:zb}
\end{align}
\begin{table}[t]
\captionstyle{center} \centering
\renewcommand{\captionfont}{\small}
\begin{tabular}{cc|cc}
\emph{subband $A$}& \emph{subband $B$} & Power & Rate (${\log}P$)\\
$x_{c,A}$ & $x_{c,B}$ & $P{-}P^\beta$ & $1{-}\beta$\\
$u_A$ & $v_B$ & $P^\alpha/2$ & $\alpha$\\
$u_0$ & $u_0$ & $(P^\beta-P^\alpha)/2$ & $\beta-\alpha$\\
$v_A$ & $u_B$ & $P^\beta/2$ & $\beta$
\end{tabular}
\caption{Power and rate allocation in the optimal scheme.}\label{tab:power_rate}
\end{table}

From \eqref{eq:ya} to \eqref{eq:zb} (ignoring the noise terms), $x_{c{,}A}$ and $x_{c{,}B}$ are first decoded by treating all the other terms as noise. Afterwards, user 1 decodes $u_0$ and $u_A$ from $y_A$ using Successive Interference Cancelation (SIC). With the knowledge of $u_0$, $u_B$ can be obtained from $y_B$. Similarly, user 2 decodes $u_0$ and $v_B$ from $z_B$ via SIC. $v_A$ can be decoded from $z_A$ by eliminating $u_0$. The sum \emph{DoF} therefore is $d_\Sigma^{opt}{=}1{+}\frac{\beta{+}\alpha}{2}$.

\subsection{Unmatched Case: Shortness of the Scheme Proposed in \cite{icc13freq}}\label{shortness}
The transmit signals in \emph{subband $A$} and $B$ are expressed as
\begin{align}
\mathbf{s}_A &{=}\left[x_{c,A}{,}0\right]^T\!\!\!{+}\left[\mu_A{,}0\right]^T\!\!\!{+}
[\hat{\mathbf{h}}_A^\bot{,}\hat{\mathbf{h}}_A][v_{A1}{,}v_{A2}]^T{+}\hat{\mathbf{g}}_A^\bot{u_A},\label{eq:sA2}\\
\mathbf{s}_B &{=}\left[x_{c,B}{,}0\right]^T\!\!\!{+}\left[\mu_B,{0}\right]^T\!\!\!{+}
[\hat{\mathbf{g}}_B^\bot{,}\hat{\mathbf{g}}_B][u_{B1}{,}u_{B2}]^T{+}\hat{\mathbf{h}}_B^\bot{v_B},\label{eq:sB2}
\end{align}
where the private symbols $u_A$, $v_{A1}$, $u_{B1}$ and $v_B$ are precoded and transmitted with the power and rate similar to $u_A$, $v_A$, $u_B$ and $v_B$ in \eqref{eq:sA} and \eqref{eq:sB} respectively.

Besides, $v_{A2}$ and $u_{B2}$, generated with rate $(\beta{-}\alpha){\log}P$ similar to $u_0$, are respectively overheard by user 1 in \emph{subband $A$} and by user 2 in \emph{subband $B$}, thus leading to the requirement of transmitting $\mu{=}v_{A2}{+}u_{B2}$ to enable the decoding of other private symbols. $\mu$ is split into $\mu_A$ and $\mu_B$ and multicast via an extra $\beta{-}\alpha$ channel use. However, in the optimal scheme, $u_0$ is the only symbol causing interference at receiver 2 in \emph{subband $A$} and is simply removed after retransmission in \emph{subband $B$}.

To sum up, the scheme in \cite{icc13freq} employs $2\beta{+}\beta{-}\alpha$ channel use to transmit six private symbols (i.e. $v_{A1}$, $v_{A2}$, $u_A$, $u_{B1}$, $u_{B2}$, $v_B$), while the optimal scheme sends five private symbols (i.e. $u_A$, $v_B$, $u_0$, $v_A$, $u_B$) in $2\beta$ channel use. Their sum \emph{DoF} are respectively expressed as (regardless of $x_{c{,}A}$ and $x_{c{,}B}$)
\begin{align}
d_\Sigma^{sub}&{=}\frac{2\beta{+}2\alpha{+}2(\beta{-}\alpha)}{3\beta{-}\alpha}=
\frac{4\alpha{+}4(\beta{-}\alpha)}{2\alpha{+}3(\beta{-}\alpha)},\label{eq:dsub}\\
d_\Sigma^{opt}&{=}\frac{2\beta{+}2\alpha{+}(\beta{-}\alpha)}{2\beta}=
\frac{4\alpha{+}3(\beta{-}\alpha)}{2\alpha{+}2(\beta{-}\alpha)}.\label{eq:dopt}
\end{align}

\eqref{eq:dsub} and \eqref{eq:dopt} provide an explicit interpretation of how the channel resources have been used. More precisely, we can see the optimal scheme is an integration of ZFBF and $S_3^{3/2}$ in \cite{Tandon12} while the scheme in \cite{icc13freq} combines ZFBF with MAT. Specifically, similarly to the weighted sum in \eqref{eq:w_sum}, both schemes employ $\alpha$ channel use to achieve the region $\bar{\mathcal{D}}$. $\bar{\mathcal{D}}$ corresponds to the optimal region with perfect CSIT of both users and can be simply achieved by ZFBF. However, over the $\beta{-}\alpha$ channel use where the CSIT state alternates between \emph{subchannel $\hat{A}$} and $\hat{B}$, the optimal scheme achieves $1.5(\beta{-}\alpha){\log}P$ sum rate, consistent with $\hat{\mathcal{D}}$, outperforming the scheme in \cite{icc13freq} (with $\frac{4}{3}(\beta{-}\alpha){\log}P$). Hence, the shortness of the scheme in \cite{icc13freq} lies in the sub-optimality of MAT in \emph{subchannel $\hat{A}$} and $\hat{B}$.

\subsection{Matched Case}\label{match_ach}

The region shown in Figure \ref{fig:Dm} can be achieved by transmitting the signals in each subband as
\begin{equation}
\mathbf{s}_i{=}[\underbrace{x_{c,i}}_{P{-}P^j}{,}0]^T{+}\underbrace{\hat{\mathbf{g}}_i^\bot{u_i}}_{P^j/2}
{+}\underbrace{\hat{\mathbf{h}}_i^\bot{v_i}}_{P^j/2}, (i{,}j){=}(A{,}\beta){,}(B{,}\alpha).\label{eq:si}
\end{equation}
$x_{c{,}i}$ is decoded first at each user with rate $(1{-}j){\log}P$ in \emph{subband $i$}. Afterwards, due to partial ZFBF, the private symbols $u_i$ and $v_i$ can be respectively decoded with rate $j{\log}P$ at user 1 and user 2.

\section{Switching among Sub-optimal Strategies}\label{switch}
As the optimal schemes discussed in Section \ref{unmatch_ach} and \ref{match_ach} operate as an integration of FDMA, ZFBF and $S_{3}^{3/2}$ (in unmatched case), we will evaluate the sum \emph{DoF} performance of sub-optimal (and less complex) transmission strategies.

\subsection{Sub-optimal Strategies}
\subsubsection{FDMA}
The sum \emph{DoF} is $d_\Sigma^F{=}1$, resulted from simply sending $x_{c{,}i}$ ($i{=}A{,}B$) in \eqref{eq:sA}, \eqref{eq:sB} and \eqref{eq:si} with full power.

\subsubsection{ZFBF}
The transmission boils down to ZFBF if only $v_i$ and $u_i$ ($i{=}A{,}B$) are sent with full power $P$. The received signal are respectively $y_i{=}\mathbf{h}_i^H\hat{\mathbf{g}}_i^\bot{u_i}{+}\mathbf{h}_i^H\hat{\mathbf{h}}_i^\bot{v_i}{+}\epsilon_1$ and $z_i{=}\mathbf{g}_i^H\hat{\mathbf{h}}_i^\bot{v_i}{+}\mathbf{g}_i^H\hat{\mathbf{g}}_i^\bot{u_i}{+}\epsilon_2$ at user 1 and 2, where the second terms are residual interferences with covariance $P\mathcal{E}[||\tilde{\mathbf{h}}_i||^2]$ and $P\mathcal{E}[||\tilde{\mathbf{g}}_i||^2]$, causing rate loss. Hence, for both unmatched and matched CSIT scenario, the sum \emph{DoF} is $d_\Sigma^Z{=}\beta{+}\alpha$.

\subsubsection{$S_3^{3/2}$}
For the unmatched case, if only $u_0$, $v_A$ and $u_B$ are transmitted using the full power, the transmission boils down to $S^{3/2}_3$ in \cite{Tandon12}. Since the qualities of $\hat{\mathbf{h}}_A^\bot$ and $\hat{\mathbf{g}}_B^\bot$ are $\beta$, $u_0$ is decoded with rate $\beta{\log}P$ by user 1 from $y_A$ and user 2 from $z_B$. As a consequence, $u_B$ and $v_A$ are respectively obtained with rate ${\log}P$ from $y_B$ and $z_A$. Hence, the sum \emph{DoF} is $d_\Sigma^{S}{=}1{+}\frac{\beta}{2}$.

\subsection{Numerical Results}
\begin{figure}[t]
\renewcommand{\captionfont}{\small}
\captionstyle{center}
\centering \subfigure[$\rho{=}90\%$]{
                \centering
                \includegraphics[width=0.22\textwidth,height=2.5cm]{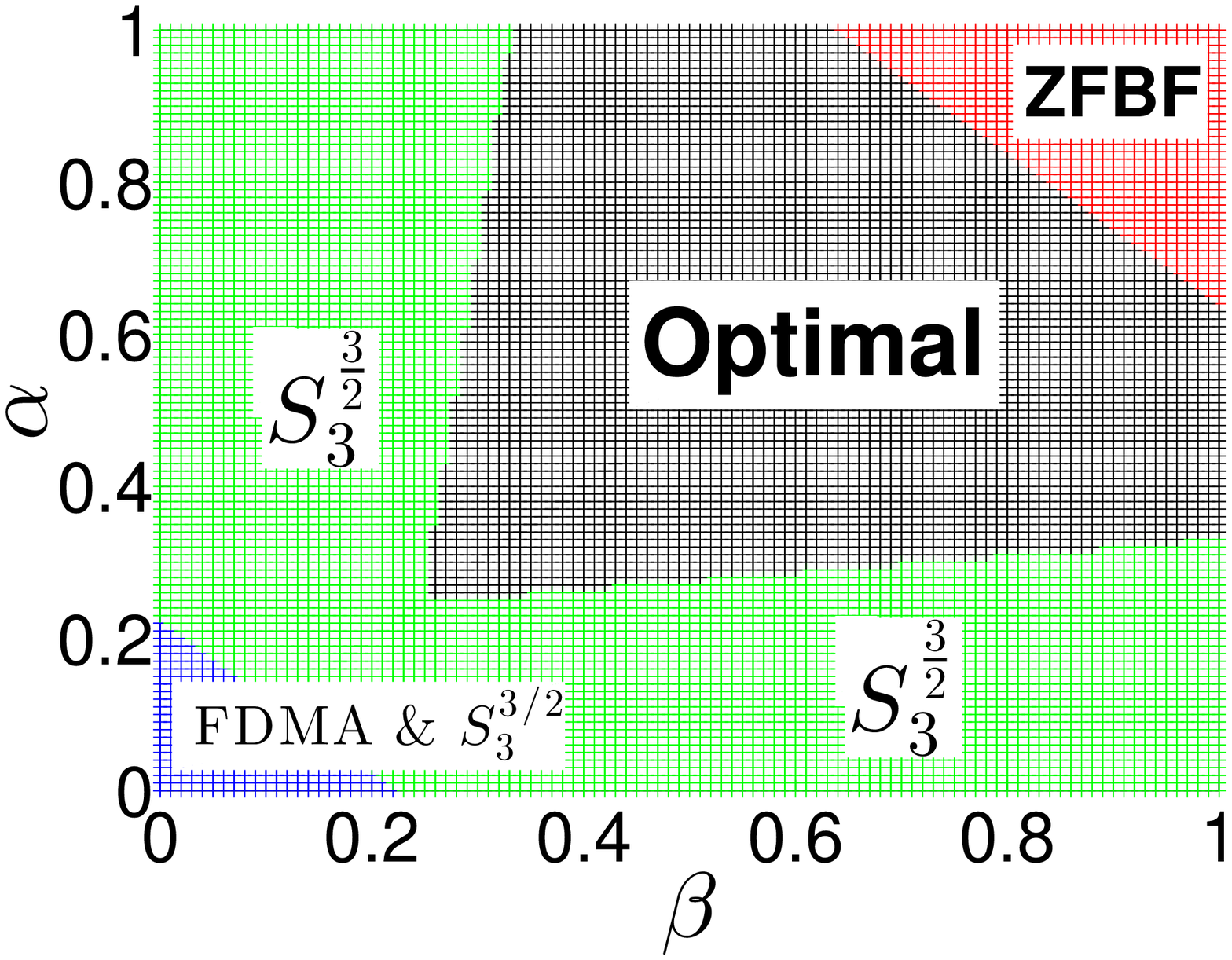}
                \label{fig:ab90}
        }
\subfigure[$\rho{=}80\%$]{
                \centering
                \includegraphics[width=0.22\textwidth,height=2.5cm]{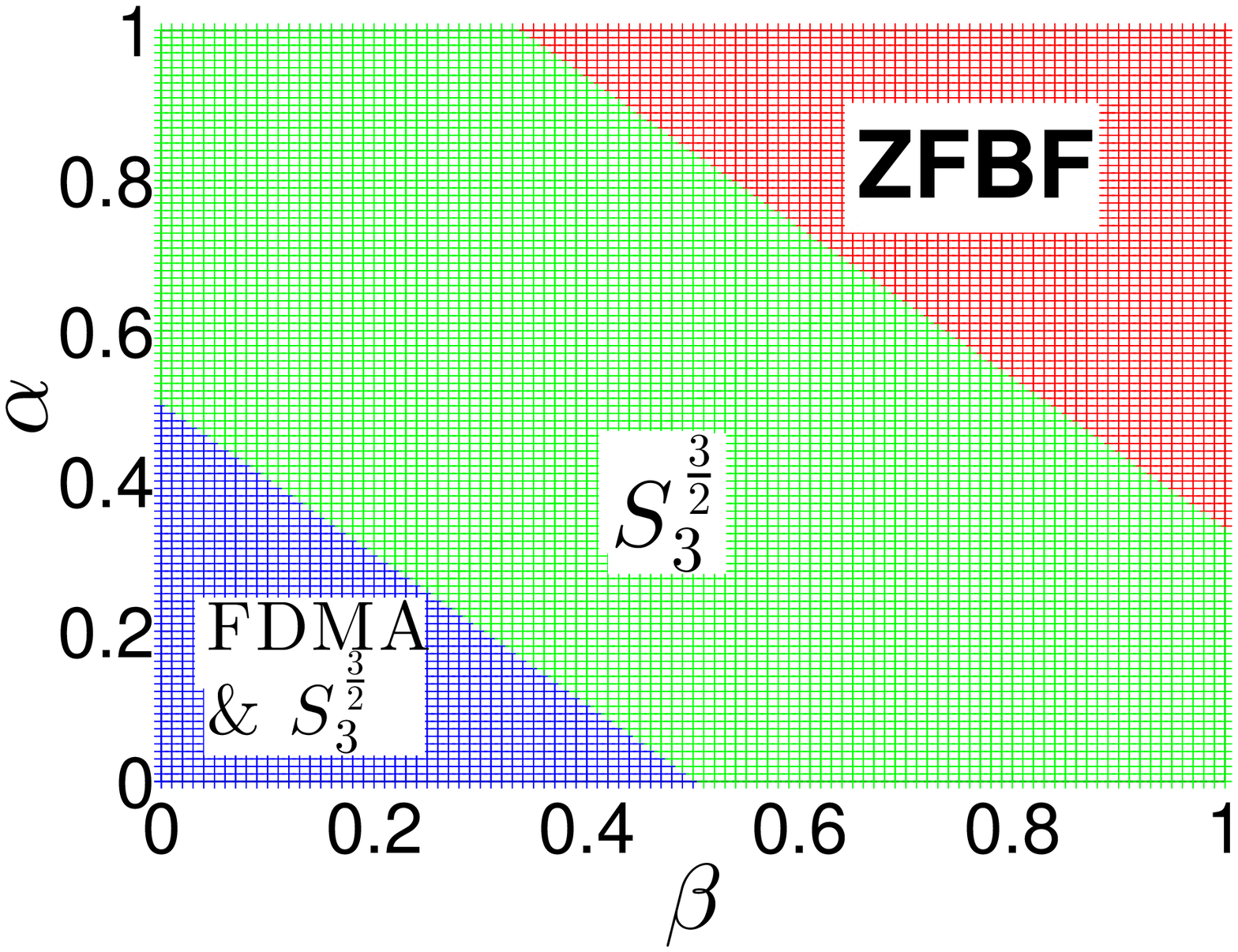}
                \label{fig:ab80}
        }
\caption{Unmatched case, switching among FDMA, ZFBF and $S^{3/2}_3$.}\label{fig:ab}
\end{figure}

\begin{figure}[t]
\renewcommand{\captionfont}{\small}
\captionstyle{center}
\centering \subfigure[$\rho{=}75\%$]{
                \centering
                \includegraphics[width=0.225\textwidth,height=2.5cm]{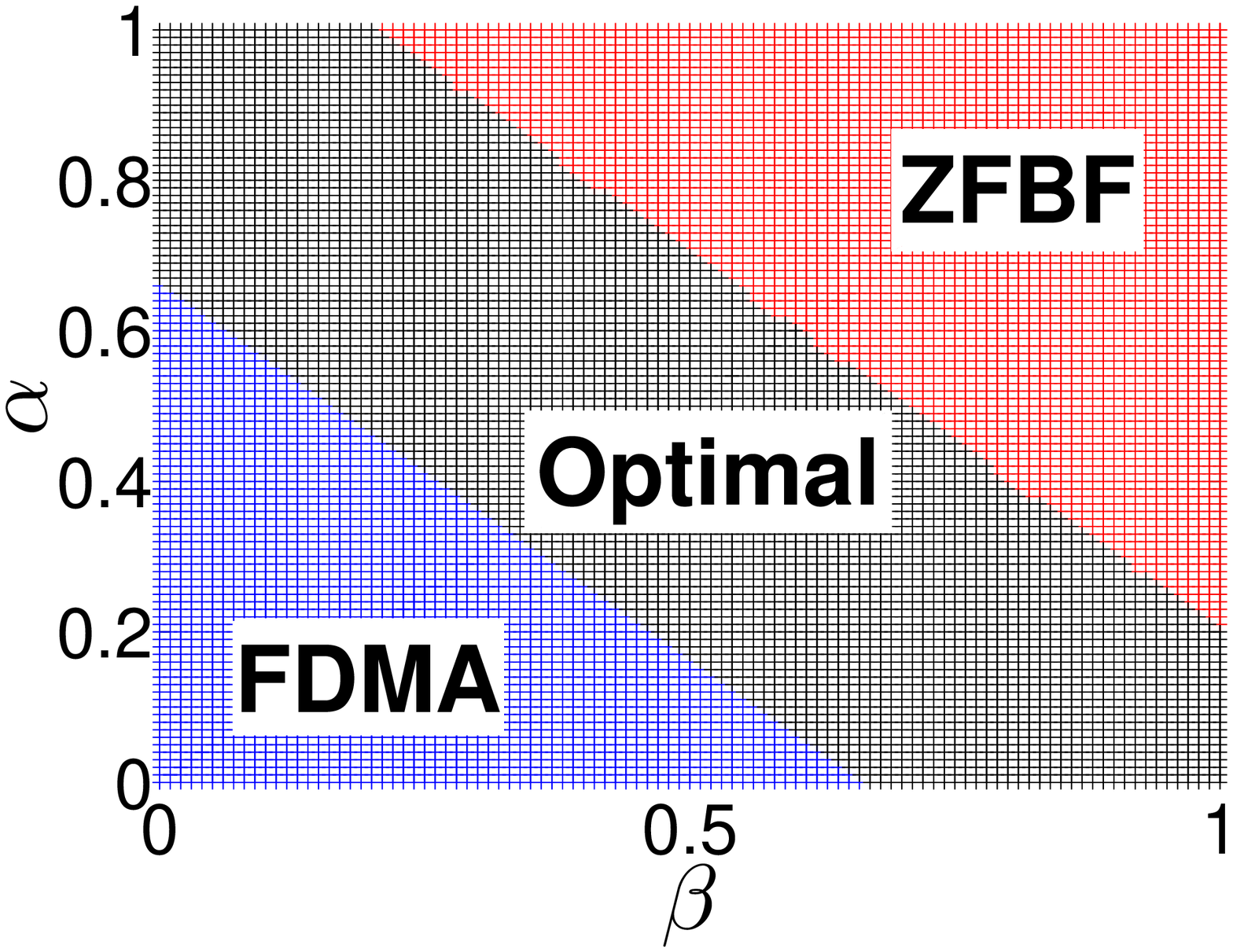}
                \label{fig:ab75_m}
        }
\subfigure[$\rho{=}66.7\%$]{
                \centering
                \includegraphics[width=0.225\textwidth,height=2.5cm]{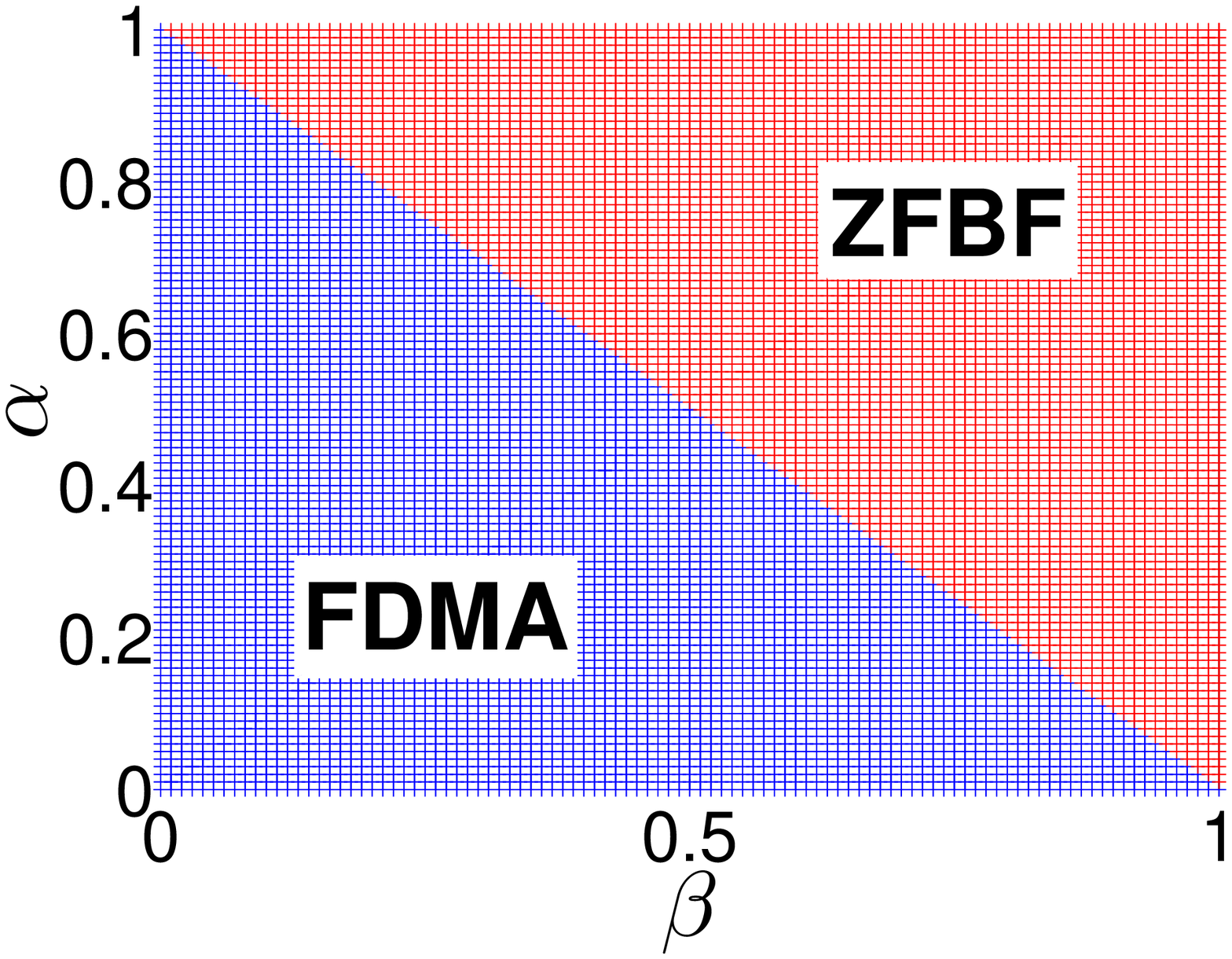}
                \label{fig:ab67_m}
        }
\caption{Matched case, switching among FDMA and ZFBF.}\label{fig:ab_m}
\end{figure}

Next, for all possible values of $\beta$ and $\alpha$, we take the max sum \emph{DoF} performance over the aforementioned sub-optimal strategies. If the max sum \emph{DoF} can achieve at least $\rho$ (expressed in $\%$) of the optimal result, the complicated optimal strategy is replaced by the sub-optimal one. Figure \ref{fig:ab} and \ref{fig:ab_m} illustrate the selection results for the unmatched and matched case respectively.

In Figure \ref{fig:ab90}, nearly an half of the $(\beta{,}\alpha)$-grid is covered by the optimal scheme when $\rho{=}90\%$. ZFBF has distinguished performance when $(\beta{,}\alpha)$ approach $1$, because the CSIT works well in rejecting the interference potentially overheard by users. $S_3^{3/2}$ scheme occupies the corners where $\beta$ and $\alpha$ have relatively large discrepancy, as one user's rate is significantly limited in each subband if ZFBF is conducted. Both FDMA and $S_3^{3/2}$ can achieve above $90\%$ of the optimal sum \emph{DoF} performance when $\beta{+}\alpha{\leq}0.2$.

Figure \ref{fig:ab80} displays an interesting result, namely that the best transmission strategy out of three covers all the possible pairs of $(\beta{,}\alpha)$ when the target is decreased to $80\%$. In other words, the best strategy among the 3 sub-optimal strategies can achieve at least $80\%$ of the optimal sum \emph{DoF} performance as
\begin{equation}
\max(d_\Sigma^F,d_\Sigma^Z,d_\Sigma^S)\geq0.8\times d_\Sigma^{opt}, \forall \beta,\alpha\in[0,1].\label{eq:p80}
\end{equation}
\eqref{eq:p80} can be derived by thoroughly comparing $d_\Sigma^F$, $d_\Sigma^Z$ and $d_\Sigma^S$ for different values of $(\beta{,}\alpha)$. For the matched case, a similar observation results from Figure \ref{fig:ab_m} as
\begin{equation}
\max(d_\Sigma^F,d_\Sigma^Z)\geq2/3\times d_\Sigma^{opt}, \forall \beta,\alpha\in[0,1].\label{eq:p67_m}
\end{equation}

\section{Conclusion}\label{conclusions}

In this contribution, we derive the outer-bound of the \emph{DoF} region in (unmatched and matched) frequency correlated scenario introduced in \cite{icc13freq}, thus showing the optimality of the achievable \emph{DoF} bound found in \cite{Elia13}. The bound is interpreted as a weighted sum of the \emph{DoF} bound achieved by FDMA, ZFBF and $S_3^{3/2}$ in \cite{Tandon12}. The origin of the sub-optimality of the scheme in \cite{icc13freq} is clarified.

We have evaluated the sum \emph{DoF} performance of simple sub-optimal transmission schemes (FDMA, ZFBF or $S_3^{3/2}$) for specific values of $(\beta{,}\alpha)$. The results show that for the unmatched CSIT scenario, the optimal scheme proposed in \cite{Elia13} can be avoided if we aim at achieving $80\%$ of the optimal sum \emph{DoF} performance. In the matched CSIT scenario, the optimal scheme can be replaced by FDMA or ZFBF provided that the level of achievement is lower than $66.7\%$.

\section*{Acknowledgement}

We acknowledge fruitful discussions on the Proof of Theorem \ref{outerbound} with our colleague Borzoo Rassouli.

\section*{Appendix-Proof of \eqref{eq:lemma1}}

To obtain \eqref{eq:lemma1}, firstly we introduce
\begin{align}
\Phi_j{=}&h(Z_j^{n{-}j{+}1}|\Omega{,}Y_1^{j{-}1}{,}Z_{n{-}j{+}2}^n)\nonumber\\
&{-}h(Y_{j}^{n{-}j{+}1}|\Omega{,}Y_1^{j{-}1}{,}Z_{n{-}j{+}2}^n), \text{for } j{\leq}\lfloor\frac{n{+}1}{2}\rfloor,\label{eq:phi}\\
\Theta_j{=}&h(Z_j|\Omega{,}Y_1^{j{-}1}{,}Z_{j{+}1}^n){-}h(Y_j|\Omega{,}Y_1^{j{-}1}{,}Z_{j{+}1}^n).\label{eq:theta}
\end{align}
The last two terms in \eqref{eq:lemma1} can be rewritten as
\begin{equation}
\Phi_1{=}\sum_{j{=}1}^n\Theta_{j}.\label{eq:lemma1_1}
\end{equation}
\eqref{eq:lemma1_1} can be obtained by summing the following recursive formulas and removing the identical terms
\begin{align}
\Phi_j{=}&\Theta_j{+}\Theta_{n{-}j{+}1}{+}\Phi_{j{+}1}, \text{for } j{\leq}\lfloor\frac{n{+}1}{2}\rfloor{-}1,\label{eq:recursive}
\end{align}
\begin{align}
\Phi_{\lfloor\frac{n{+}1}{2}\rfloor}{=}&\left\{\begin{array}{ll}\Theta_{\lfloor\frac{n{+}1}{2}\rfloor}
{+}\Theta_{\lfloor\frac{n{+}1}{2}\rfloor{+}1},&\text{if n is even,}\\ \Theta_{\lfloor\frac{n{+}1}{2}\rfloor},&\text{if n is odd.}\end{array}\right.
\end{align}

\eqref{eq:recursive} originates from the following derivations
\begin{align}
&h(Z_j^{n{-}j{+}1}|\Omega{,}Y_1^{j{-}1}{,}Z_{n{-}j{+}2}^n){-}h(Y_{j}^{n{-}j{+}1}|\Omega{,}Y_1^{j{-}1}{,}Z_{n{-}j{+}2}^n)\nonumber\\
{=}&h(Z_j|\Omega{,}Y_1^{j{-}1}{,}Z_{j{+}1}^n){+}h(Z_{j{+}1}^{n{-}j{+}1}|\Omega{,}Y_1^{j{-}1}{,}Z_{n{-}j{+}2}^n)\nonumber\\
&{-}h(Y_{n{-}j{+}1}|\Omega{,}Y_1^{n{-}j}{,}Z_{n{-}j{+}2}^n){-}h(Y_{j}^{n{-}j}|\Omega{,}Y_1^{j{-}1}{,}Z_{n{-}j{+}2}^n)\nonumber\\
{=}&h(Z_j|\Omega{,}Y_1^{j{-}1}{,}Z_{j{+}1}^n){+}h(Z_{j{+}1}^{n{-}j{+}1}|\Omega{,}Y_1^{n{-}j}{,}Z_{n{-}j{+}2}^n)\nonumber\\
&{+}I(Z_{j{+}1}^{n{-}j{+}1};Y_{j}^{n{-}j}|\Omega{,}Y_1^{j{-}1}{,}Z_{n{-}j{+}2}^n)\nonumber\\
&{-}h(Y_{n{-}j{+}1}|\Omega{,}Y_1^{n{-}j}{,}Z_{n{-}j{+}2}^n){-}h(Y_{j}^{n{-}j}|\Omega{,}Y_1^{j{-}1}{,}Z_{j{+}1}^n)\nonumber\\
&{-}I(Y_{j}^{n{-}j};Z_{j{+}1}^{n{-}j{+}1}|\Omega{,}Y_1^{j{-}1}{,}Z_{n{-}j{+}2}^n)\nonumber\\
{=}&h(Z_j|\Omega{,}Y_1^{j{-}1}{,}Z_{j{+}1}^n){+}h(Z_{n{-}j{+}1}|\Omega{,}Y_1^{n{-}j}{,}Z_{n{-}j{+}2}^n)\nonumber\\
&{+}h(Z_{j{+}1}^{n{-}j}|\Omega{,}Y_1^{n{-}j}{,}Z_{n{-}j{+}1}^n)\nonumber\\
&{-}h(Y_{n{-}j{+}1}|\Omega{,}Y_1^{n{-}j}{,}Z_{n{-}j{+}2}^n){-}h(Y_{j}|\Omega{,}Y_1^{j{-}1}{,}Z_{j{+}1}^n)\nonumber\\
&{-}h(Y_{j{+}1}^{n{-}j}|\Omega{,}Y_1^{j}{,}Z_{j{+}1}^n)\nonumber\\
{=}&\Theta_j{+}\Theta_{n{-}j{+}1}\nonumber\\
&{+}h(Z_{j{+}1}^{n{-}j}|\Omega{,}Y_1^{j}{,}Z_{n{-}j{+}1}^n){-}I(Z_{j{+}1}^{n{-}j};Y_{j{+}1}^{n{-}j}|\Omega{,}Y_1^{j}{,}Z_{n{-}j{+}1}^n)\nonumber\\
&{-}h(Y_{j{+}1}^{n{-}j}|\Omega{,}Y_1^{j}{,}Z_{n{-}j{+}1}^n){+}I(Y_{j{+}1}^{n{-}j};Z_{j{+}1}^{n{-}j}|\Omega{,}Y_1^{j}{,}Z_{n{-}j{+}1}^n)\nonumber\\
{=}&\Theta_j{+}\Theta_{n{-}j{+}1}{+}\Phi_{j{+}1}.
\end{align}

Note that we learn that \eqref{eq:lemma1_1} is known as Csisz\'ar Sum Identity \cite{network_info} after we work out the derivation.

\bibliographystyle{IEEEtran}

\bibliography{pimrc2013}

\end{document}